\documentclass[12pt,a4paper,dvips]{article}

\usepackage{graphicx}
\usepackage{amsbsy}
\usepackage{amssymb}

\begin{document}

\newcommand{\be}{\begin{equation}}
\newcommand{\ee}{\end{equation}}
\newcommand{\bea}{\begin{eqnarray}}
\newcommand{\eea}{\end{eqnarray}}
\newcommand{\tr}{\,\hbox{tr }}
\newcommand{\Tr}{\,\hbox{Tr }}
\newcommand{\Det}{\,\hbox{Det }}
\newcommand{\fslash}{\hskip-0.22cm /}
\newcommand{\Fslash}{\hskip-0.26cm /\ }
\begin{titlepage}
\vspace*{1.5cm}
\begin{center}
{\Large \bf  Abelian Bosonization, the Wess-Zumino Functional and
Conformal Symmetry\\}
\vspace*{1cm}
Aleksandar Bogojevi\' c, Branislav Sazdovi\' c\\
{\it Institute of Physics\\
      P.O.Box 57, Belgrade 11001, Yugoslavia\\}
\vspace*{5mm}
Olivera Mi\v skovi\' c\\
{\it Institute of Nuclear Sciences ``Vin\v ca"\\
      Department for Theoretical Physics\\
      P.O.Box 522, Belgrade 11001, Yugoslavia\\}
\vspace*{5mm}
Preprint IP-HET-98/14\\
October 1998\\
\end{center}

\vspace*{1cm}
\begin{abstract}
We look at the equivalence of the massive Thirring and sine-Gordon models. 
Previously, this equivalence was derived perturbatively in mass (though to 
all orders). Our calculation goes beyond that and uncovers an underlying conformal symmetry. 
\end{abstract}
\end{titlepage}

\section{Introduction}

Field theories in two dimensions display many interesting properties that have no direct analogy in higher dimensions. The most characteristic such property  is related to the spin of a particle. In two dimensions the rotation group is Abelian and spin is a continuous parameter. For this reason there can exist an equivalence between a bosonic and a fermionic theory. The realization of this idea is called bosonization. The roots of bosonization can be traced to the work of Klaiber \cite{klaiber} who looked at the massive Thirring model \cite{thirring}, as well as of Lowenstein and Swieca \cite{ls} and Schwinger \cite{schwinger} who looked at the massless Schwinger model. They analysed certain non-trivial aspects of the corresponding theories, but stoped short of bosonizing them. Bosonization was first carried through by Coleman \cite{coleman} and Mandelstam \cite{mandelstam}. They both worked in the operator formalism, as did a host of authors that followed them and that focused on equalities of the corresponding current algebras and energy momentum tensors. Derivations in the functional formalism followed after the seminal paper by Fujikawa \cite{fujikawa}. All the derivations follow through for the massless Thirring model. However, in the case of the massive Thirring model the best results so far have given a perturbative (in powers of mass) derivation of the equivalence (although to all orders).

In this paper we work with the massive Thirring model in the functional formalism. With the help of standard Wess-Zumino techniques we construct an equivalent gauge theory, and analyze two possible gauge fixings. One gauge fixing, by construction, leads to the massive Thiring model. The other gives us the sine-Gordon model. In its general outlines, this idea is related to the so-called continuous bosonization of Damgaard, et al. \cite{damgaard}. Unlike them, we use the Wess-Zumino technique to construct our wider gauge theory. The central point of our derivation lies in the location of a conformal symmetry and in its gauge fixing.

\section{Gauge Anomalies}

In this section we give a brief review of how to deal with theories
with gauge anomalies. Let us look at a quantum field theory with matter
fields $\Phi$ and gauge fields $A$. The partition function equals
\be
Z=\int {\cal D}\Phi\,{\cal D}A\,
e^{-\frac{1}{\hbar}I[A,\Phi]}=
\int {\cal D}A\,e^{-\frac{1}{\hbar}W[A]}\ .
\ee
The above theory has a gauge anomaly if the action is invariant under a
given local transformation while the measure is not. To be specific,
we assume that
\bea
I[A^U,\Phi^U ] &=& I[A,\Phi ]\nonumber\\
{\cal D}\Phi^U &=& {\cal D}\Phi\,e^{\alpha[A;U]}\nonumber\\
{\cal D}A^U    &=& {\cal D}A\ .
\eea
Using the above we easily see that the anomaly $\alpha$ satisfies
\be
W[A^U]-W[A]=\hbar\,\alpha [A;U]\ .
\ee
By executing two gauge transformations $U$ and $U'$ one after the other we find that the anomaly satisfies the cocycle condition
\be
\alpha [A^U;U']-\alpha [A;UU']+\alpha [A;U]=0\ .\label{cocycle}
\ee
We can now modify our starting theory in a way that gets rid of the
gauge anomaly. To do this we change the measure according to
\be
{\cal D}\Phi\to \overline{{\cal D}\Phi}=
{\cal D}\Phi\,{\cal D}H\,e^{\alpha[A;H]}\ ,
\ee
where $H$ is a new field that takes its values in the gauge group, and
${\cal D}H$ is the the appropriate Haar measure satisfying 
${\cal D}H={\cal D}H^U$. The modified theory is gauge invariant if we have
\be
\alpha [A^U;H^U]-\alpha [A;H]+\alpha [A;U]=0\ .
\ee
Because of the cocycle condition (\ref{cocycle}) the above equation is
satisfied if the auxilliary field $H$ transforms according to
\be
H^U=U^{-1}H\ .
\ee
In this way we have obtained a new theory
\be
e^{-\frac{1}{\hbar}\overline W[A]}=\int {\cal D}\Phi\,{\cal D}H\,
e^{-\frac{1}{\hbar}\bar I[A,\Phi ,H]}\ ,
\ee
given in terms of the new action
\be
\bar I[A,\Phi ,H]=I[A,\Phi ]-\hbar\,\alpha [A;H]\ .
\ee
As we can see, $I$ and $\bar I$ are classically equivalent. Note that
$\bar I$ corresponds to an effective theory --- it has explicit
$\hbar$ dependence, and is in general non-local because of the Wess-Zumino
term $\alpha [A;H]$. This is the price we pay for getting rid of the
anomaly. Note also that by gauge fixing the new theory with $H=1$ we
recover our starting model. In the rest of the paper we will set $\hbar=1$.

\section{The Massive Thirring Model}

The massive Thirring model is a model of fermions in $d=2$ dimensions
with Lagrangian
\be
{\cal L}_\mathrm{MTM}=\bar\psi(\partial\fslash+m)\psi+
\frac{1}{2}\,gj^2\ ,
\ee
where $j_\mu=\bar\psi\gamma_\mu\psi$. The quartic interaction can be
simplified by introducing an additional vector field $A_\mu$, so that
\be
{\cal L}_\mathrm{MTM}=\bar\psi(D\Fslash+m)\psi-
\frac{1}{2g}\,A_\mu^2\ .
\ee
The covariant derivative is defined to be $D_\mu=\partial_\mu+A_\mu$.
We write the partition function as
\be
Z_\mathrm{MTM}=\int {\cal D}\mu\,e^{-\int dx\,{\cal L}_0}\ ,
\ee
where ${\cal L}_0=\bar\psi D\Fslash\psi$, while the measure is
given by
\be
{\cal D}\mu={\cal D}\psi {\cal D}\bar\psi {\cal D}A\,
e^{-\int dx\,\left(m\,\bar\psi\psi-\frac{1}{2g}\,A_\mu^2\right)}\ .
\label{measure}
\ee
${\cal L}_0$ is invariant under local vector transformations
\bea
\psi &\to& \psi^\omega=e^{i\omega}\,\psi\nonumber\\  
\bar\psi &\to& \bar\psi^\omega=e^{-i\omega}\,\bar\psi\nonumber\\
A_\mu &\to& A_\mu^\omega=A_\mu-i\partial_\mu\omega\ .
\eea
The measure ${\cal D}\mu$ is not invariant. Because of the additional
exponential term in (\ref{measure}) we find
\be
{\cal D}\mu\to {\cal D}\mu^{\omega}={\cal D}\mu\,
e^{-\int dx\,\left(\frac{1}{2g}\,(\partial\omega)^2+
\frac{i}{g}\,A\cdot\partial\omega\right)}\ .
\ee
At the same time, ${\cal L}_0$ is invariant under local axial vector
transformations
\bea
\psi &\to& \psi^\lambda=e^{i\lambda\gamma_5}\,\psi\nonumber\\  
\bar\psi &\to& \bar\psi^\lambda=\bar\psi e^{i\lambda\gamma_5}\nonumber\\
A_\mu &\to& A_\mu^\lambda=A_\mu+\varepsilon_{\mu\nu}
\partial_\nu\lambda\ .
\eea
Again, ${\cal D}\mu$ is not invariant. The non invariance comes from the
exponential term in (\ref{measure}), but also from the well known axial
anomaly of ${\cal D}\psi{\cal D}\bar\psi$. Taken together we find
\bea
\lefteqn{
{\cal D}\mu\to {\cal D}\mu^{\lambda}=}\nonumber\\
& &{}={\cal D}\mu\,
e^{\int dx\,\left(\frac{1}{2g}\,(\partial\lambda)^2+
\frac{1}{g}\,\varepsilon_{\mu\nu}A_\mu\partial_\nu\lambda
-m\,\bar\psi\psi(\cos 2\lambda - 1)-
im\,\bar\psi\gamma_5\psi\sin 2\lambda\right)}\cdot\nonumber\\
& &{}\qquad\cdot e^{\,\frac{1}{2\pi} \int dx\,\left( (\partial\lambda)^2+
2\varepsilon_{\mu\nu}A_\mu\partial_\nu\lambda\right)}\ .
\eea
The first term comes about from the exponential term in the definition of
${\cal D}\mu$, the second from the anomaly of ${\cal D}\psi{\cal D}\bar\psi$.

Using the perscription of the previos section we obtain a new, anomaly free
theory
\be
\overline Z_\mathrm{MTM}=\int
{\cal D}\psi{\cal D}\bar\psi{\cal D}A{\cal D}\varphi{\cal D}\phi\,
e^{-\int dx\, \overline{\cal L}}\ ,
\ee
where
\bea
\overline{\cal L} &=& \bar\psi\partial\fslash\psi+
j\cdot A-\frac{1}{2g}\,A^2+
\frac{1}{2g}(\partial\varphi)^2+\frac{i}{g}\,A\cdot\partial\varphi+
\nonumber\\
&&+m\,\bar\psi\psi\cos 2\phi+im\,\bar\psi\gamma_5\psi\sin 2\phi-
\nonumber \\
&&-\frac{1}{2}\left(\frac{1}{g}+\frac{1}{\pi}\right)(\partial\phi)^2
-\left(\frac{1}{g}+\frac{1}{\pi}\right)\varepsilon_{\mu\nu}
A_\mu\partial_\nu\phi\ .
\eea
Note that $\varphi$ represents the auxilliary field coming from the vector
symmetry {\it i.e.} $H_V=e^{i\varphi}$. Similarly, $\phi$ is the auxilliary
field due to the axial vector symmetry, so $H_A=e^{i\gamma_5\phi}$. Because
of this, under vector transformations we have $\varphi\to\varphi-\omega$,
$\phi\to\phi$. On the other hand, under axial vector transformations
$\varphi\to\varphi$, $\phi\to\phi-\lambda$. Integrating out the vector field
and rescaling $\phi\to \frac{\beta}{2}\,\phi$, where
$\frac{\beta^2}{4\pi}=\frac{1}{1+g/\pi}$, we find
\bea
\overline{\cal L} &=& \bar\psi\partial\fslash\psi+
\frac{1}{2}\,gj^2+
m\,\bar\psi\psi\cos\beta\phi+im\,\bar\psi\gamma_5\psi\sin\beta\phi+
\nonumber \\
&&+ ij\cdot\partial\varphi+\frac{1}{2}\,(\partial\phi)^2-
\frac{2\pi}{\beta}\varepsilon_{\mu\nu}j_\mu\partial_\nu\phi\ .
\label{lbar}
\eea
Gauge fixing this with $\phi=\varphi=0$ gets us back to the massive Thirring
model.

As has been recently shown \cite{thomassen}, by integrating out the fermions
we obtain, in the functional formalism, the well known result that the
massive Thirring model is equivalent to the sine-Gordon model through the
bosonization  relations
\bea
\psi_R &=& \sqrt{\frac{c\mu}{2\pi}}\,:e^{-i\beta\sigma_R}:\nonumber\\
\psi_L &=& \sqrt{\frac{c\mu}{2\pi}}\,:e^{i\beta\sigma_L}:\ ,
\eea
where $\psi=\left(\begin{array}{c}
             \psi_R\\
             \psi_L
             \end{array}\right)$. In the above bosonization relations
$c=e^\gamma$ and $\gamma$ is the Euler constant
$\gamma=\lim_{n=\to\infty}\left(\sum_{k=1}^n
\frac{1}{k}-\ln n\right)\approx 0.577$, and $\mu\to 0$ is an infra red
regulator. In terms of fermi bilinears this gives
\newpage
\bea
\bar\psi\psi &=& \frac{c\mu}{\pi}\,:\cos\beta\sigma:\nonumber\\
\bar\psi\gamma_5\psi &=& -i\frac{c\mu}{\pi}\,:\sin\beta\sigma:\nonumber\\
j_\mu &=& \frac{\beta}{2\pi}\,\varepsilon_{\mu\nu}\partial_\nu\sigma\ .
\eea
Our aim in this paper is not to integrate out the fermions in (\ref{lbar}),
but rather to find a different gauge fixing that leads to the sine-Gordon
model. To do this, let us first look at some simple
consequences of bosonization. First of all (\ref{lbar}) is
consistent with the above bosonization relations. Using them
the Lagrangian in (\ref{lbar}) is transformed into
\be
\overline{\cal L}=\frac{1}{2}\,\left(\partial_\mu(\phi-\sigma)\right)^2-
\frac{mc\mu}{\pi}\cos\beta(\phi-\sigma)\ ,
\ee
which is indeed the sine-Gordon Lagrangian. We see that it is trivialy
invariant under
\bea
\phi &\to& \phi-\lambda\nonumber\\
\sigma &\to& \sigma-\lambda\ .
\eea
Compare this to our axial symmetry.
We can gauge fix this symmetry by choosing $\sigma=0$
and obtain
\bea
\bar\psi\psi &=& \frac{c\mu}{\pi}\nonumber\\
\bar\psi\gamma_5\psi &=& 0\nonumber\\
j_\mu &=& 0\ .\label{fix}
\eea

We are ready to gauge fix (\ref{lbar}) directly
without using the bosonization relations. It is convenient to write
the $U(1)_V\times U(1)_A$ symmetry as $U(1)_R\times U(1)_L$. We find
\bea
\psi_R &\to& e^{i\theta_R}\psi_R\nonumber\\
\psi_L &\to& \psi_L\ ,
\eea
as well as
\bea
\psi_R &\to& \psi_R\nonumber\\
\psi_L &\to& e^{i\theta_L}\psi_L\ ,
\eea
where $\theta_R=\lambda+\omega$, $\theta_L=\lambda-\omega$. We thus
choose to fix the gauge by imposing $\psi_R^\dagger=\psi_R$,
$\psi_L^\dagger=\psi_L$. Using this we get the gauge fixed Lagrangian
\be
{\cal L}=2\psi_L\partial_z\psi_L+2\psi_R\partial_{\bar z}\psi_R+
\frac{1}{2}\,(\partial\phi)^2+2im\,\psi_R\psi_L\cos\beta\phi\ .
\label{l}
\ee
We have introduced complex coordinates $z=x_0+ix_1$, $\bar z=x_0-ix_1$.
Things have indeed simplified. As before in (\ref{fix}), we again have
$\bar\psi\gamma_5\psi=0$ and $j_\mu=0$. Now, however, we seem to have no
further symmetries at our disposal whose gauge fixing would make
$\bar\psi\psi=2i\psi_R\psi_L$ equal to a constant, as was the case in
(\ref{fix}), and this is precisely what we do need in order for
(\ref{l}) to give the sine-Gordon Lagrangian.

The resolution of this problem is simple, though rather interesting.
We must give a more precise treatment of the field $\phi$ coming from the
Wess-Zumino term for the axial anomaly. It is well known that there are
problems with massless scalar fields in $d=2$ dimensions. The way
around these problems is to introduce an infra red regulator $\mu$ through
the free correlator
\be
\langle\phi(x)\phi(y)\rangle =-\frac{1}{\beta^2}\ln c^2 \mu^2(x-y)^2\ .
\label{cor}
\ee
This is the source of $\mu$ and $c$ in the bosonization relations. At the
end of all calculations we take the $\mu\to 0$ limit. From (\ref{cor})
we see that we are not dealing with one field $\phi(x)$, but rather with a
family of fields $\phi(x|\mu)$ --- one for each choice of $\mu$. Written
in the operator formalism, the piece of the Lagrangian (\ref{l}) that seems
to be causing problems is
\be
m\,\bar\psi(x)\psi(x):\cos\beta\phi(x|\mu):\,=
\frac{m}{\mu}\,\bar\psi(x)\psi(x)\,\mu:\cos\beta\phi(x|\mu):\ .
\ee
Note that $\bar\psi(x)\psi(x)$ and $\mu:\cos\beta\phi(x|\mu)$ are both
conformal fields of scaling dimension $(\frac{1}{2},\frac{1}{2})$.
A field $\xi$ has conformal dimension $(a,b)$ if under
$z\to z'=f(z)$, $\bar z\to \bar z'=\bar f(\bar z)$ it transforms as
\be
\xi(z,\bar z)\to \xi'(z',\bar z')=
\left(\frac{df}{dz}\right)^{-a} \left(\frac{d\bar f}{d\bar z}\right)^{-b}
\xi(z,\bar z)\ .
\ee
The whole Lagrangian is, therefore, of scaling dimension $(1,1)$. This just
compensates the transformation of $dzd\bar z$. The only remaining constant
in the theory is $m/\mu$, which is dimensionless. As a consequence we
find that $\overline{\cal L}$ is invariant under a further symmetry ---
conformal symmetry. It is percisely this symmetry that allows us to fix
$\bar\psi\psi$ to be a constant. Under a conformal transformation
we have
\be
\bar\psi(x)\psi(x)\to
\left(\frac{df}{dz}\right)^{-1/2}
\left(\frac{d\bar f}{d\bar z}\right)^{-1/2}
\bar\psi(x)\psi(x)\ .
\ee
By taking $\psi_R=\theta\left(\frac{df}{dz}\right)^{1/2}$ and
$\psi_L=\bar\theta\left(\frac{d\bar f}{d\bar z}\right)^{1/2}$ we find that
$\bar\psi\psi$ indeed becomes a constant. Note that $\theta$ and $\bar\theta$
are Grassmann constants, while $\theta\bar\theta$ is an ordinary commuting
number.

\section{Conclusion}

In conclusion, we have re-derived the Abelian bosonization results of Coleman \cite{coleman}, Mandelstam \cite{mandelstam} and others \cite{dorn}--\cite{damgaard} concerning the equivalence between the massive Thirring and sine-Gordon models. Unlike our derivation of the equivalence, all the previous ones were perturbative (to all orders) in the mass $m$. As we have shown, the central point in our derivation is the existence of \emph{two} mass scales $m$ and $\mu$, and the fact that they enter solely through their ratio $\frac{m}{\mu}$.

\end{document}